# Optimizing Starshot lightsail design: a generative network-based approach

Zhaxylyk A. Kudyshev[1,2], Alexander V. Kildishev[1], Vladimir M. Shalaev[1,2], Alexandra Boltasseva[1,2*]

[1]School of Electrical and Computer Engineering, Birck Nanotechnology Center and Purdue Quantum Science and Engineering Institute, Purdue University, West Lafayette, IN, 47907, USA

[2]The Quantum Science Center (QSC), a National Quantum Information Science Research Center of the U.S. Department of Energy (DOE), Oak Ridge, TN, 37931, USA

KEYWORDS: Starshot Lightsail, inverse design, topology optimization, machine learning

**ABSTRACT:** The Starshot lightsail project aims to build an ultralight spacecraft ("nanocraft") that can reach Proxima Centauri b in approximately 20 years, requiring propulsion with a relativistic velocity of ~60 000 km/s. The spacecraft's acceleration approach currently under investigation is based on applying the radiation pressure from a high-power laser array located on Earth to the spacecraft lightsail. However, the practical realization of such a spacecraft imposes extreme requirements to the lightsail's optical, mechanical, thermal properties. Within this work, we apply adjoint topology optimization and variational autoencoder-assisted inverse design algorithm to develop and optimize a silicon-based lightsail design. We demonstrate that the developed framework can provide optimized optical and opto-kinematic properties of the lightsail. Furthermore, the framework opens up the pathways to realizing a multi-objective optimization of the entire lightsail propulsion system, leveraging the previously demonstrated concept of physics-driven compressed space engineering.

## INTRODUCTION

Radiation pressure – the momentum exchange between incident electromagnetic wave and the object - is an essential phenomenon for many applications, such as optical tweezers[1,2] and solar-sail-assisted space travel[3]. One of the most notable efforts to realize a lightsail space mission is IKAROS (Interplanetary Kite-craft Accelerated by Radiation of the Sun).[4,5] The IKAROS spacecraft leverages the radiation pressure from the sunlight, achieving the maximum velocity of 400 m s$^{-1}$.[6,7] The Starshot Breakthrough Initiative has recently initiated the ambitious Starshot lightsail project, which aims at building an ultralight nanocraft that can reach Proxima Centauri b in about 20 years. The Starshot mission targets the development of the space 'nanocraft' that can achieve a relativistic speed of ~60 000 km s$^{-1}$ (20% of the speed of light) utilizing the optical force from the high-power laser array located on Earth. The development of the Starshot nanocraft design is a multi-objective problem, which requires optimization over the interdisciplinary domain to meet the extreme requirements for its optical, thermal, and mechanical properties. To reach the aforementioned relativistic velocity, the lightsail should have (i) an extremely low mass density (less than 1 g per 10 m$^2$ surface area), (ii) an exceptionally high broadband reflectivity within the Doppler-shift bandwidth, and (iii) a low optical absorption coefficient (absorptivity less than 10$^{-5}$).[6] While the requirement on the absorption can be addressed by choosing the suitable material and the working wavelength of the laser source, the mass and optical properties should be optimized by minimizing the acceleration distance ($D$) that involves a tradeoff between broadband reflectivity and the total mass of the lightsail.[6] Previously, it has been demonstrated that silicon photonic crystal structures with simplified geometries could be optimized to provide $D \sim 4 \times 10^9$ m.[6] Then, the inverse design utilizing the method of moving asymptotes has been adapted to designing the lightsail with the specific goal of minimizing $D$.[8] Surprisingly, the inverse design framework that allowed optimizing the device performance with at least 10$^4$ spatial design variables converged to a simple, almost intuitive one-dimensional subwavelength grating, providing up to 50% improvement of the lightsail's optical/mass properties ($D \sim 1.9 \times 10^9$ m).[8]

Recently, machine and deep learning algorithms have emerged as powerful tools for solving inverse design problems in nanophotonics through optimization of shape/topologies of the sub-diffractive resonant photonic/plasmonic elements – metastructures.[9–16] It has been demonstrated that generative networks could be efficiently coupled to conventional optimization frameworks, such as metaheuristic and gradient descent algorithms, to address multi-objective nanophotonic problems.[17–19] Within this work, we apply two approaches to optimizing lightsail design to meet mass/reflectivity objectives. The first approach adapts traditional adjoint topology optimization (TO) to maximize the broadband reflectivity of a silicon metastructure over Doppler-shift bandwidth. The second approach leverages an autoencoder-based generative network trained on the initial set of topology optimized lightsail designs, using it for rapid sampling of the optimized free-form metastructures for the Starshot lightsail.

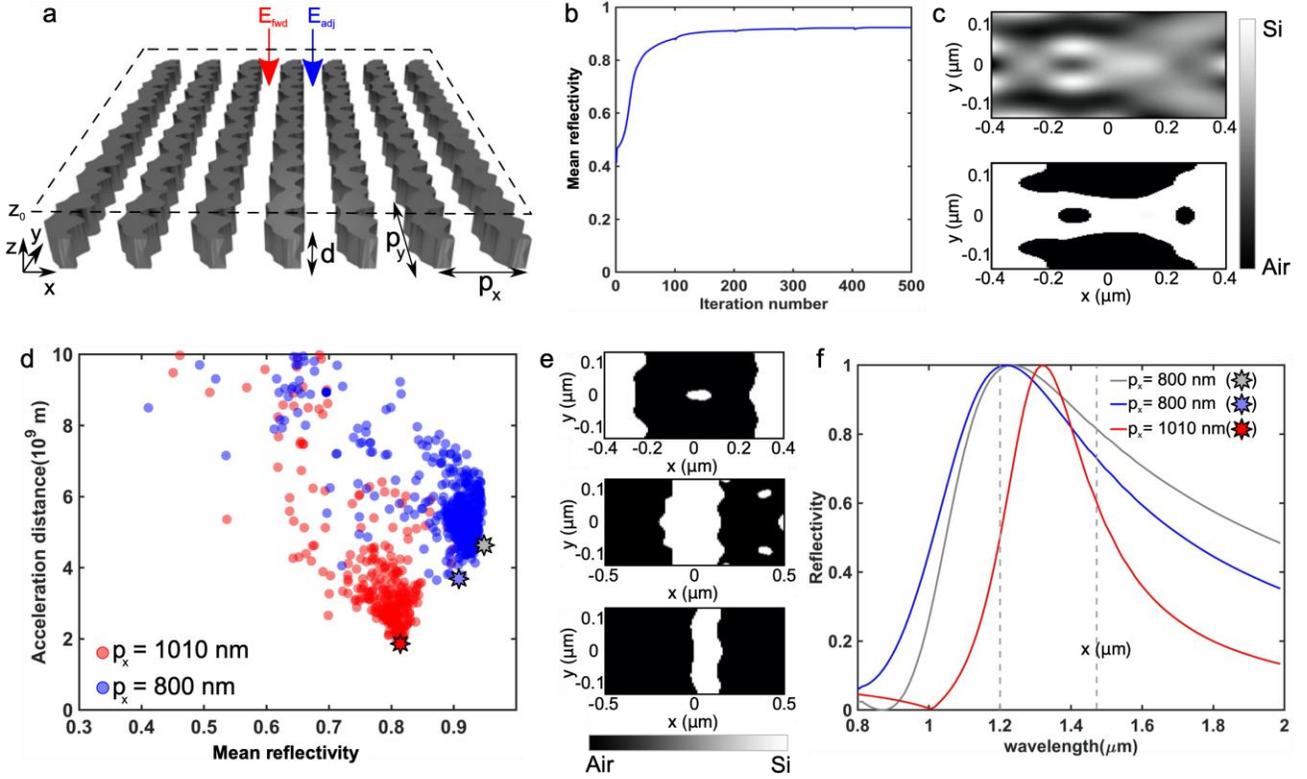

**Figure 1. Topology optimization of the lightsail design.** (a) Schematics of the lightsail design and optimization setup configuration. (b) Convergence of the mean reflectivity during the adjoint topology optimization run. (c) Material distribution within the optimization region in the beginning (top) and in the end (bottom) of the optimization run. (d) Distribution of the design performance on the acceleration distance vs. mean reflectivity map for two cases of periodicity, Case 1, $p = \{800\ nm,\ 266\ nm\}$ (blue), and Case 2, $p = \{1010\ nm,\ 336\ nm\}$ (red). The red star and blue star markers show the Case 1 and Case 2 designs with the lowest acceleration distance in the set. The gray star shows the position of the Case 1 design with highest reflectivity. (e) Corresponding unit cell configuration for gray (top), blue (middle) and red (bottom) star markers. (f) Reflectivity spectra of lightsail designs shown in (e). Dashed vertical lines show the corresponding Doppler-shift bandwidth.

## TOPOLOGY OPTIMIZATION OF LIGHTSAIL DESIGN

The ideal lightsail, due to its high reflectivity, should enable effective propulsion of the nanocraft via momentum transfer from incident laser radiation at wavelength $\lambda_0$. The critical point to consider is the Doppler shift of the photons of the incident laser ($\lambda_f = 1.22\lambda_0$), arising due to the lightsail acceleration to its target velocity ($v_f = 0.2c_0$). Hence, the first optimization objective is to maximize the lightsail reflectivity within the Doppler-shift bandwidth $\lambda \in [\lambda_0, 1.22\lambda_0]$. Another objective of the optimization is to minimize the total mass of the nanocraft (sail and payload), which is essential from the point of view of the energy/size requirement for the Earth-based laser array. The latter objective can be met by designing a lightsail with a minimal acceleration distance ($D$) – a total distance required by the nanocraft to reach $v_f$. $D$ accounts for the tradeoff between the optical (reflectivity) and kinematic (mass) properties of the lightsail and can be approximately assessed as:[20]

$$D = \frac{m_\Sigma c_0}{2 I_0 S} \int_0^{v_f} \frac{v\, dv}{\left(1 - \frac{v}{c_0}\right)^2 \sqrt{1 - \left(\frac{v}{c_0}\right)^2}} \quad (1)$$

here $m_\Sigma = m_{sail} + m_{payload}$ is the total mass of the nanocraft, $I_0$ is the laser intensity, and $S$ is the lightsail surface area, and $c_0$ is the free-space speed of light. Integration in (1) is done over the velocity (v) range from zero to the target velocity $v_f$.

Within this work, we adapt an adjoint TO[21–25] to maximizing the reflectivity of free-form silicon metastructures within the Doppler-shift bandwidth. Once the TO run is executed, the performance of each design is assessed with corresponding values of $D$.

In more detail, the TO aims at maximizing the overlap integral between the field distributions calculated just above the optimization region ($z = z_0$) (Figure 1a) during the forward $[E^{fwd}, H^{fwd}]$ and adjoint $[\bar{E}^{adj}, \bar{H}^{adj}]$ simulation runs. This integral defines the figure of merit (FoM) function, formally written as,[26]

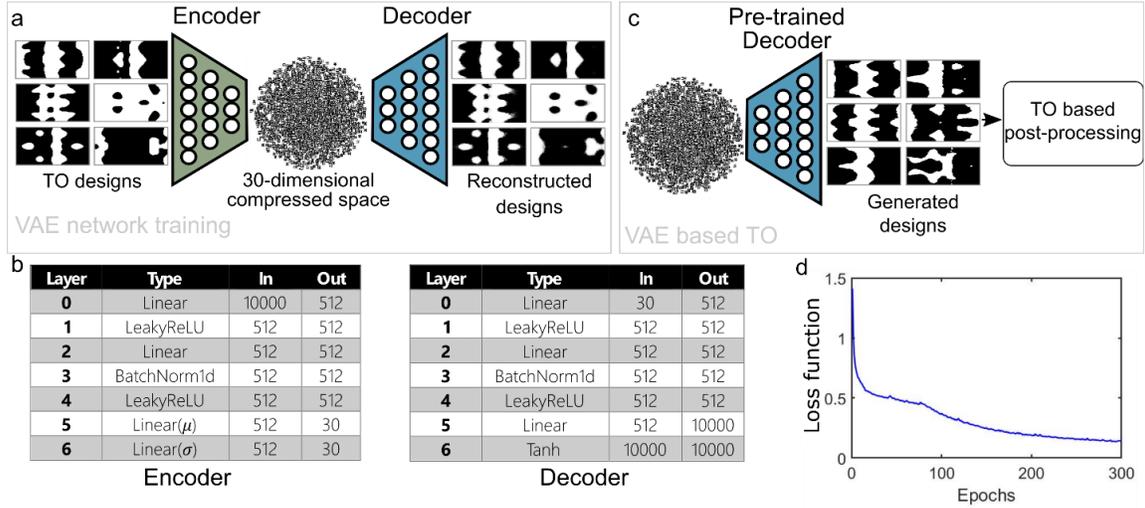

**Figure 2. VAE-TO optimization scheme.** (a) Training of the variational autoencoder (VAE) on topology optimized lightsail designs. Insets show the input and reconstructed TO lightsail designs. (b) Description of the encoder and the decoder structure. Table contains: layer number, type, sizes of in and out features of each layers. (c) VAE based topology optimization. Inset shows examples of generated TO designs. (d) Evolution of the loss function $\mathcal{L}_{VAE}$ during the training.

$$FoM(\lambda) = \left| \int_{x_0}^{x_0+p_x} \int_{y_0}^{y_0+p_y} \begin{pmatrix} \boldsymbol{E}^{\text{fwd}}(x,y,z_0;\lambda) \times \overline{\mathbf{H}}^{\text{adj}}(x,y,z_0;\lambda) - \\ -\overline{\boldsymbol{E}}^{\text{adj}}(x,y,z_0;\lambda) \times \boldsymbol{H}^{\text{fwd}}(x,y,z_0;\lambda) \end{pmatrix} \cdot \boldsymbol{n}_z dy dx \right|^2 \quad (2)$$

here $p_x, p_y$ are lateral dimensions of the device unit cell. The main task of the TO is to determine the combination of binary material distribution $\varepsilon(x,y)$ that maximizes (2). Discretizing the unit cell into $N_x \times N_y$ optimization grid, a brute-force approach would require $2^{N_x \times N_y}$ permutations to calculate the gradient of FOM at each iteration. In contrast, the adjoint TO formalism allows to significantly reduce the number of simulation runs needed to determine the spatial distribution of the FoM gradient down to two per iteration (forward and adjoint). This reduction of the computational cost is achieved by applying the Born approximation, representing the first-order variance of the permittivity function within each pixel as a set of independent (induced) point dipoles.[25] This approximation, in turn, allows calculating the sensitivity of the $FoM(\lambda)$ function to any possible $\delta\varepsilon$ perturbation via the product of the field inside the structure induced by the dipole sources, corresponding Green tensors, and the field from adjoint simulation. Applying reciprocity theorem to the Green tensors, the final form of the spatial distribution of the wavelength-dependent gradient of FoM is expressed as a product of the E-fields inside the structure, obtained from the forward and adjoint simulation steps:

$$\left.\frac{\partial FOM}{\partial \varepsilon}\right|_{x,y,\lambda} = 2\varepsilon_0 \omega^2 \Delta V \Re\left[\bar{r}(\lambda)\boldsymbol{E}^{\text{fwd}}\left(x,y,\frac{d}{2},\lambda\right) \cdot \boldsymbol{E}^{\text{adj}}\left(x,y,\frac{d}{2};\lambda\right)\right] \quad (3)$$

here $\bar{r}$ is a conjugated wavelength-dependent complex reflection coefficient, $\Delta V$ is the volume of the grid cell of the optimization region, $\varepsilon_0$ is the free-space permittivity, $\omega = \frac{2\pi c_0}{\lambda}$ is the frequency of the incident light, and $d$ is the thickness of metastructure. Finally, to optimize the broadband response of the lightsail, we calculate the total gradient as a weighted sum over all discrete wavelength values of (3) as,

$$\left.\frac{\partial \text{FoM}}{\partial \varepsilon}\right|_{x,y} = \sum_{i=1}^{N_\lambda}(1-|\boldsymbol{r}(\lambda_i)|^2)\left.\frac{\partial \text{FoM}}{\partial \varepsilon}\right|_{x,y,\lambda_i}$$

where $N_\lambda$ is the number of wavelengths taken within the Doppler-shift bandwidth.

We optimize a sub-diffractive silicon metastructure with a thickness of 110 nm and two cases of periodicity: Case 1, $\boldsymbol{p} = \{800 \text{ nm}, 266 \text{ nm}\}$ and Case 2, $\boldsymbol{p} = \{1010 \text{ nm}, 336 \text{ nm}\}$. Setting the working wavelength of the incident laser to $\lambda_0 = 1.2$ μm gives the Doppler-shift range, $\lambda \in [1.2$ μm, $1.464$ μm]. Figure 1b shows the convergence of reflectivity averaged over the Doppler-shift bandwidth during the TO run. The optimization region is discretized into a 100×100 optimization grid. Initially, the material distribution in the optimization region is set to be a random, smooth dielectric function (Figure 1c, top). At the same time, with the progress of the TO run, it converges into a binary material distribution (Figure 1c, bottom) after applying the binarization function each 50th iteration. The filtering procedures eliminate sub-25-nm features, while robustness control also applied during the TO ensures that the final design is insensitive to fabrication perturbations. Rigorous coupled-wave analysis (RCWA) is used for forward and adjoint simulations.[27]

In both cases, we optimized 800 lightsail designs. Performance of an optimized design is assessed with corresponding acceleration distance $D$. We consider the realistic values of the payload mass ($m_{\text{payload}} = 10$ g), the total area of lightsail (S = 10 m²), the intensity ($I_0 = 10 \frac{GW}{m^2}$), and wavelength ($\lambda_0 = 1.2$ μm) of the incident laser beam. Figure 1d shows the acceleration distance vs. mean reflectivity map of the optimized designs for Cases 1 and 2. The figure indicates that Case 1 provides a much higher mean reflectivity of the lightsail design; however, a larger filling factor of the silicon inside the unit cell leads to moderate values of $D$. The design with the highest reflectivity ($R = 0.95$) and lowest acceleration distance ($D = 4.59 \times 10^9\ m$) is shown with the green marker in Figure 1d. The blue marker is used for the design with the lowest acceleration distance ($D = 3.68 \times 10^9 m$) and a reflectance of $R = 0.91$.

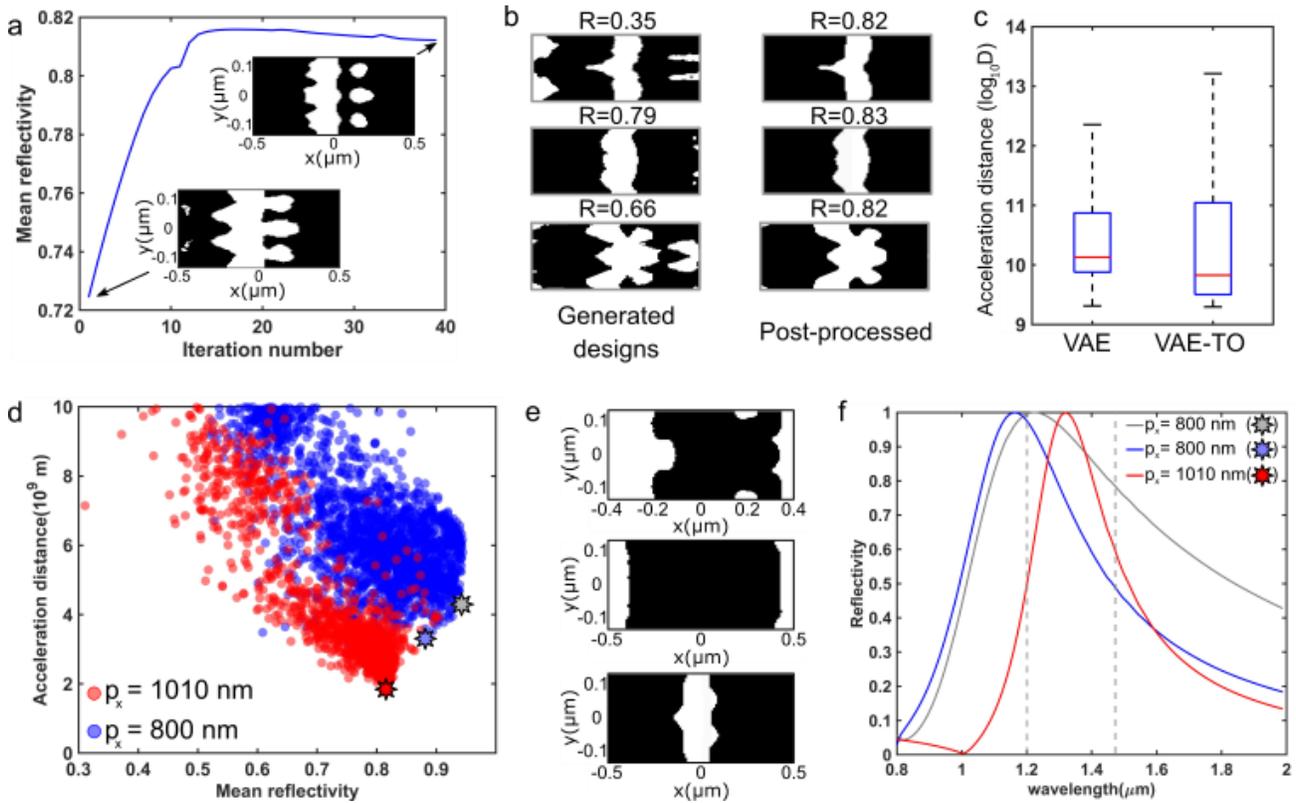

**Figure 3. VAE-TO optimization results of lightsail design.** (a) Convergence plot of mean reflectivity during TO post-processing. Insets show initial(bottom) and final(top) material distribution. (b) Some of the unit cell configuration generated by VAE(left) and corresponding lightsail designs after TO post-processing (right). Labels show mean reflectivity values. (c) Comparison of VAE and VAE-TO distributions of acceleration distance values for VAE and VAE-TO design sets. Acceleration distance axis is in logarithmic scale. Box plot shows the median (red line), interquartile range (box), and min/max values (whiskers). (a)-(c) correspond to the Case 2. (d) Distribution of the design performance on the acceleration distance vs mean reflectivity map for two cases of periodicity: Case 1, $p = [800\ nm,\ 266\ nm]$ (blue), and Case 2, $p = [1010\ nm,\ 336\ nm]$ (red). The red star and blue star markers show the Case 1 and Case 2 designs with lowest acceleration distance in the set, while the gray marker shows the position of Case 1 design with the highest reflectivity. (e) Corresponding unit cell configuration for gray (top), blue (middle), and red (bottom) star markers. (f) Reflectivity spectra of the lightsail designs shown in panel (e). Vertical dashed lines indicate the Doppler-shift bandwidth.

In contrast, Case 2 provides much lower acceleration distances due to a significantly lower filling factor. However, the mean reflectivity of Case 2 is only ~0.8. The best design within the optimized set reaches the acceleration distance, as small as $D = 1.9 \times 10^9\ m$ with a reflectance of $R = 0.81$ (red marker). Corresponding material distributions inside the unit cells are shown in Figure 1e. The top figure corresponds to Case 1, gray marker; the middle unit cell configuration shows the design of Case 1 (blue marker). The lower unit cell corresponds to Case 2 (red marker). The figure demonstrates that the adjoint TO converges to a simple 1D grating type structure, similar to the method of moving asymptotes that aims at minimization of $D$ directly.[8] Figure 1f shows corresponding reflectivity spectra of optimized lightsail designs. The figure confirms that the designs show similar spectral distributions of the reflection coefficient in all cases, which is the result of general topological similarity within the unit cell (all of them are just variations of 1D silicon gratings).

## GENERATIVE NETWORK-BASED APPROACH FOR LIGHTSAIL DESIGN OPTIMIZATION

The second approach that we apply to the lightsail design is a TO hybrid scheme with variational autoencoder (VAE). This framework consists of two steps (i) construction of the compressed space of the optimization problem via training of the VAE network on TO lightsail designs, and (ii) TO of the lightsail designs based on initial conditions sampled by a pre-trained network. The VAE network includes two coupled neural networks, an encoder and a decoder (see, Figure 2a).[28] The encoder is a network with an input layer of 100×100 dimensions, a hidden dense layer containing 512 neurons with ReLU activation function, and a batch normalization layer. The decoder has an inverted structure to the encoder, two layers with 512 neurons, and an output layer. The structure of both networks is shown in Figure 2b. Specifically, the VAE network learns how to compress real space designs ($x$) into a 30-dimensional compressed space ($z$) and then reconstruct them ($\hat{x}$). The VAE is trained by minimizing both the reconstruction loss of a design ($\mathcal{L}(x,\hat{x})$) and the Kullback-Leibler divergence loss,[28]

$$\mathcal{L}_{VAE} = \mathcal{L}(x,\hat{x}) + KL[q(z|x)|p(z)] \quad (4)$$

While the reconstruction loss determines the ability of the network to reconstruct the original design correctly, the latter defines the deviation of the recognition distribution (obtained with the model data) from the pre-defined prior. The priors, $p(z)$, and the recognition model, $q(z|x)$, are assumed to be under the normal distribution. The main challenge during the training of the VAE is that the derivatives cannot be directly calculated on the stochastic nodes to properly backpropagate the error for the training of the network. The sampling of a latent variable with Gaussian distribution is realized by re-parameterization $z \sim N(\mu,\sigma)$ as $z = \mu + \sigma \cdot \varepsilon$, where $\varepsilon \sim N(0,1)$ and $(\mu,\sigma)$ are parameters of the encoder. This re-parameterization allows to calculate their derivatives with respect to $\mu$ and $\sigma$ and use $\varepsilon$ as an additional input parameter sampled during each training epoch.[29]

Figure 2d shows the convergence of the $\mathcal{L}_{VAE}$ during the training process. The figure indicates that training of the VAE network on the TO designs leads to low loss function values at the end of the training. This result indicates that the VAE network is learning the vital geometrical features of the TO lightsail designs and can construct the correct compressed space representation for the problem. Some examples of the input and reconstructed designs by the trained VAE are shown in Figure 2a, while the examples of the sampled designs by the trained decoder are shown in Figure 2c.

Previous studies demonstrated that trained generative networks could be efficiently coupled with TO via sampling antenna designs and then used as the initial conditions for TO post-processing.[17,18] Such a hybrid approach allows (i) converging to local extrema much faster than employing TO with randomly selected initial material distribution and (ii) filtering stable, high-efficiency designs out of large datasets sampled by a neural network. Here, we apply a similar approach, where the decoder of the pre-trained VAE network is used as a generator of the initial conditions for an additional TO post-processing run (Figure 2c).

We sample 3000 lightsail designs for each periodicity case and then use them as the initial conditions for TO post-processing. Figure 3a shows the convergence of the mean reflectivity as a function of TO iterations, while the initial (bottom) and final (top) material distributions are shown in the inset. The figure confirms that the TO post-processing converges within ~15 iterations due to good initial condition, while the TO with a randomly selected initial condition typically requires ~150 iterations to converge. The filtering and robustness control are applied at each 10[th] iteration, while a maximum number of iterations is set to 40. Additionally, it can be seen that due to filtering, the TO-based post-processing allows eliminating sub 25 nm features from the final design. Such features occur in the generated lightsail antennas due to unavoidable noise introduced by the VAE network. Some of the generated and final lightsail design configurations are shown in Figure 3b. The labels show mean reflectivity values before and after TO post-processing. Figure 3c compares the performances of the lightsail designs for Case 2 sampled by VAE network and after TO post-processing. The box plot shows the distribution of acceleration distances of the design set generated by VAE and the results of the VAE-TO framework. The acceleration distance axis is in the logarithmic scale. The box plot shows the median (red line), interquartile range (box, 25th to 75th percentile), and the min/max of the distributions(whiskers). The VAE sampled designs set has narrower distribution compared to VAE-TO designs with the median at $D = 1.34 \times 10^{10}$ m and the interquartile range between $7.41 \times 10^9$ m and $7.41 \times 10^{10}$ m, while the best design in the set ensures $D = 2 \times 10^9$ m. The VAE-TO design set has a $D = 6.6 \times 10^9$ m median and interquartile range between $3.62 \times 10^9$ m and $1.09 \times 10^{11}$ m. The best design in the VAE-TO set has a $D = 1.90 \times 10^9$ m. Figure 3d shows the VAE-TO distributions within the acceleration distance vs. mean reflectivity map for both cases of periodicity. Similar to the direct TO results, VAE-TO with Case 1 periodicity produces higher reflectivity values; however, due to a higher filling ratio of the unit cell, it provides relatively large values of the acceleration distance. The best design in the set gives $D = 3.4 \times 10^9$ m and $R = 0.88$ (blue star marker), while the design with the best tradeoff between acceleration distance and reflectivity ensures $D = 4.31 \times 10^9$ m and $R = 0.94$ (gray star marker). The best design in the set of Case 2 periodicity ensures $D = 1.9 \times 10^9$ m and $R = 0.81$ (red star marker), which is identical to the direct TO results. Figure 3e shows corresponding design configurations. Notably, both of the approaches, direct TO and VAE-TO, converge to almost identical lightsail design configurations due to similar reflectivity spectra of the optimized lightsail designs (Figure 3f).

## CONCLUSION

Within this work, we implemented two different inverse design frameworks for optimizing lightsail design for the Starshot space exploration mission. We applied the adjoint TO and VAE-TO hybrid frameworks to optimize two distinct cases of bi-periodic silicon metastructures that can satisfy optical and weight constraints for the lightsail nanocraft. When the acceleration distance is used as the main criterion, the inverse optimization algorithms of both frameworks converge to a simple, intuitive one-dimensional photonic grating with the acceleration distance of $D = 1.9 \times 10^9$ m and mean reflectivity around 0.81. Notably, this result is identical to the design obtained with the method of moving asymptotes, which targeted optimization of acceleration distance directly.[8] Such a simple solution to the problem results from extreme mass constrain, which demands a low filling ratio of the material inside the unit cell. Moreover, we demonstrate that the lightsail design with extremely high broadband reflectivity within the Doppler-shift bandwidth, along with moderate acceleration distances, can be obtained as well. We envision that taking into account additional optimization constraints, such as stability and/or thermal properties of the lightsail, will inevitably deviate the optimal solution from one-dimensional grating type structures towards a more complex topology, which should provide a balance between optical, kinematic, mechanical, and thermal properties of the lightsail. Such optimization will require multi-objective optimization over enlarged optimization spaces. Previously, it has been demonstrated that autoencoder-assisted inverse design frameworks can be efficiently applied to multi-objective optimization problems. These approaches could be used for global optimization via hybridization with different heuristics algorithms,[12,30] including frameworks that allow sampling globally optimized solutions by leveraging classical and quantum quadratic unconstrained binary optimization solvers.[31] The developed VAE-TO framework opens the way to realizing the multi-objective optimization of the lightsail design employing the previously proven concept of physics-driven compressed space engineering.[19]

## AUTHOR INFORMATION

**Corresponding Author**
*Alexandra Boltasseva, aeb@purdue.edu

**Author Contributions**
Z.A.K. conceived the main framework of the adjoint TO and VAE-TO. A.V.K., V.M.S, and A.B. supervised the project. All authors interpreted the data and wrote the manuscript.

## ACKNOWLEDGMENT

This work is supported by the Breakthrough Initiatives, a division of the Breakthrough Prize Foundation, and Purdue's Elmore ECE Emerging Frontiers Center.